\documentclass{article}

\usepackage[preprint]{neurips_2024}

\usepackage[utf8]{inputenc} 
\usepackage[T1]{fontenc}    
\usepackage{hyperref}       
\usepackage{url}            
\usepackage{booktabs}       
\usepackage{amsfonts}       
\usepackage{nicefrac}       
\usepackage{microtype}      
\usepackage{booktabs}
\usepackage{adjustbox}
\usepackage{array}
\usepackage{subcaption}
\usepackage{amsmath}
\usepackage{amssymb}

\usepackage{multirow}
\usepackage[table]{xcolor} 
\usepackage{cleveref}

\definecolor{xiaomiblue}{HTML}{4A7BCE}      
\definecolor{xiaomipaleblue}{HTML}{B8DCFE}  
\definecolor{xiaomiorange}{HTML}{FFA903}    
\definecolor{xiaomiteal}{HTML}{03CCA0}      
\definecolor{xiaomigreen}{HTML}{50B341}     
\definecolor{xiaomicoral}{HTML}{ED696D}     
\definecolor{xiaomilightgray}{HTML}{AAAAA8} 
\definecolor{xiaomibrightblue}{HTML}{04A3FD}
\definecolor{xiaomimedgray}{HTML}{6E6E6C}   
\definecolor{xiaomiblack}{HTML}{030303}     
\definecolor{xiaomired}{HTML}{ee4028}

\newcommand{\cmark}{\textcolor{xiaomigreen}{\checkmark}}
\newcommand{\smark}{\ensuremath{\textcolor{xiaomiblack}{\mathbf{\sim}}}}
\newcommand{\xmark}{\ensuremath{\textcolor{xiaomired}{\mathbf{\times}}}}

\title{DashengTokenizer: One layer is enough for unified audio understanding and generation}

\author{%
  Heinrich Dinkel \quad Xingwei Sun \quad Gang Li \quad Jiahao Mei \quad Yadong Niu \\ 
  \vspace{0.2cm} \\ 
  \textbf{Jizhong Liu} \quad \textbf{Xiyang Li} \quad \textbf{Yifan Liao} \quad \textbf{Jiahao Zhou} \quad \textbf{Junbo Zhang} \quad \textbf{Jian Luan}\\
  \vspace{0.2cm} \\ 
MiLM Plus, Xiaomi Inc., Beijing, China\\
  \texttt{\{dinkelheinrich,zhangjunbo1\}}@xiaomi.com 
}
\definecolor{salmon}{HTML}{FFA07A}
\definecolor{highlightblue}{HTML}{E1F5FE}
\definecolor{baselinegray}{HTML}{F2F2F2}

\begin{document}

\maketitle

\begin{abstract}

This paper introduces DashengTokenizer, a continuous audio tokenizer engineered for joint use in both understanding and generation tasks. 
Unlike conventional approaches, which train acoustic tokenizers and subsequently integrate frozen semantic knowledge, our method inverts this paradigm: we leverage frozen semantic features and inject acoustic information.
In linear evaluation across 22 diverse tasks, our method outperforms previous audio codec and audio encoder baselines by a significant margin while maintaining competitive audio reconstruction quality.
Notably, we demonstrate that this acoustic injection improves performance for tasks such as speech emotion recognition, music understanding, and acoustic scene classification. 
We further evaluate the tokenizer's generative performance on text-to-audio (TTA), text-to-music (TTM), and speech enhancement (SE).
Our approach surpasses standard variational autoencoder (VAE)-based methods on TTA and TTM tasks, while its effectiveness on SE underscores its capabilities as a general-purpose audio encoder.
Finally, our results challenge the prevailing assumption that VAE-based architectures are a prerequisite for audio synthesis.
Checkpoints are available \href{https://huggingface.co/mispeech/dashengtokenizer}{\includegraphics[height=1em,keepaspectratio]{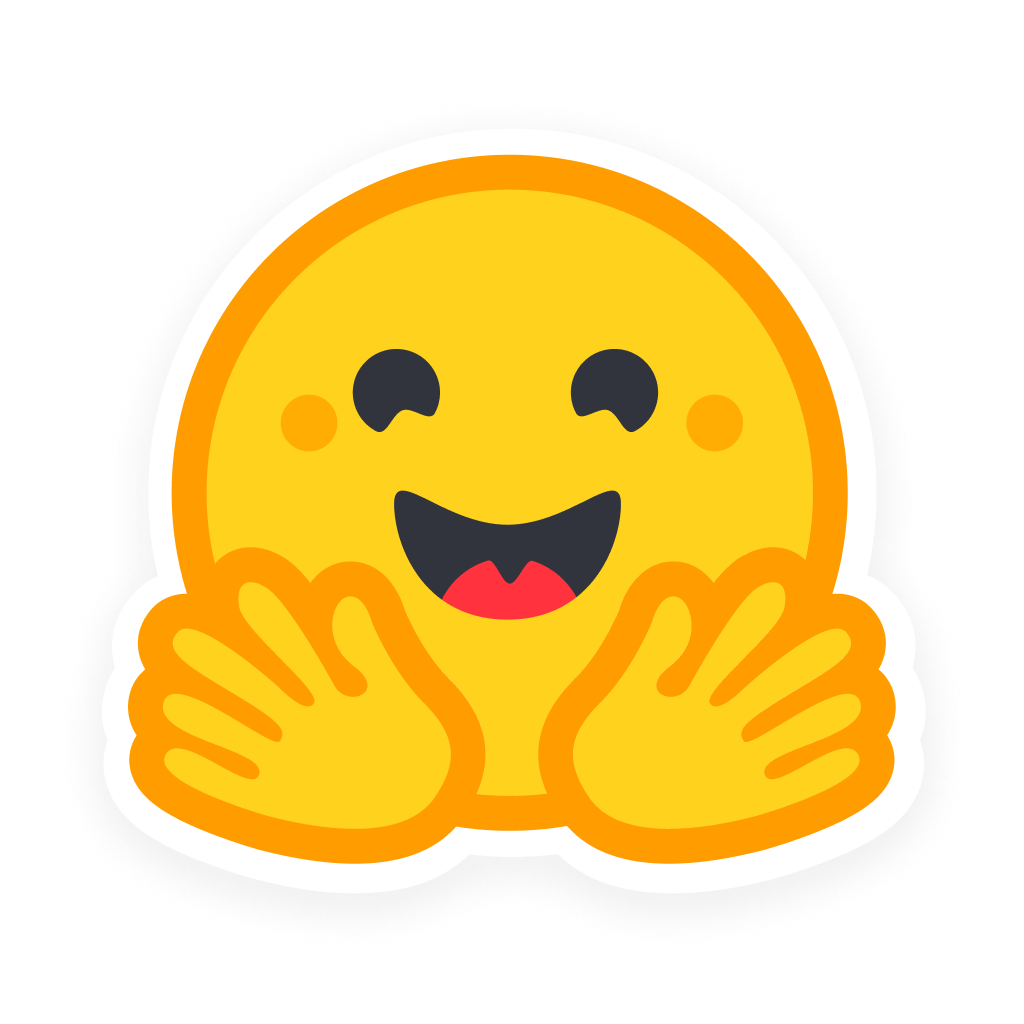}}.

\end{abstract}

\section{Introduction}

The recent surge in Generative AI has been propelled by advances in Large Language Models (LLMs) and Diffusion Models, significantly enhancing the capabilities of audio foundation models. 

Both of these models are in theory capable to be used for audio understanding~\cite{chu2024qwen2,zhou2025diffa} and audio generation~\cite{uniaudio_yang,huang2023make}, a representation gap persists in practice. 
Understanding tasks typically rely on unidirectional encoders that produce coarse, high-dimensional semantic embeddings.

In contrast, generation employs tokenizers ( discrete or continuous autoencoders ) to ensure high-fidelty reconstruction from low-dimensional acoustic features.

Current literature for joint understanding and generation thus adopt one of two architectures: (I) Employing a semantic encoder alongside an independent acoustic tokenizer. While effective, this approach is computationally redundant and increases system complexity; or (II) Training a single model to capture both semantic and acoustic information. These models frequently prioritize reconstruction, often resulting in subpar semantic representations compared to dedicated encoders

\begin{figure}[htb]
     \centering
     \begin{subfigure}[c]{0.50\linewidth}
         \centering
         \includegraphics[width=\linewidth]{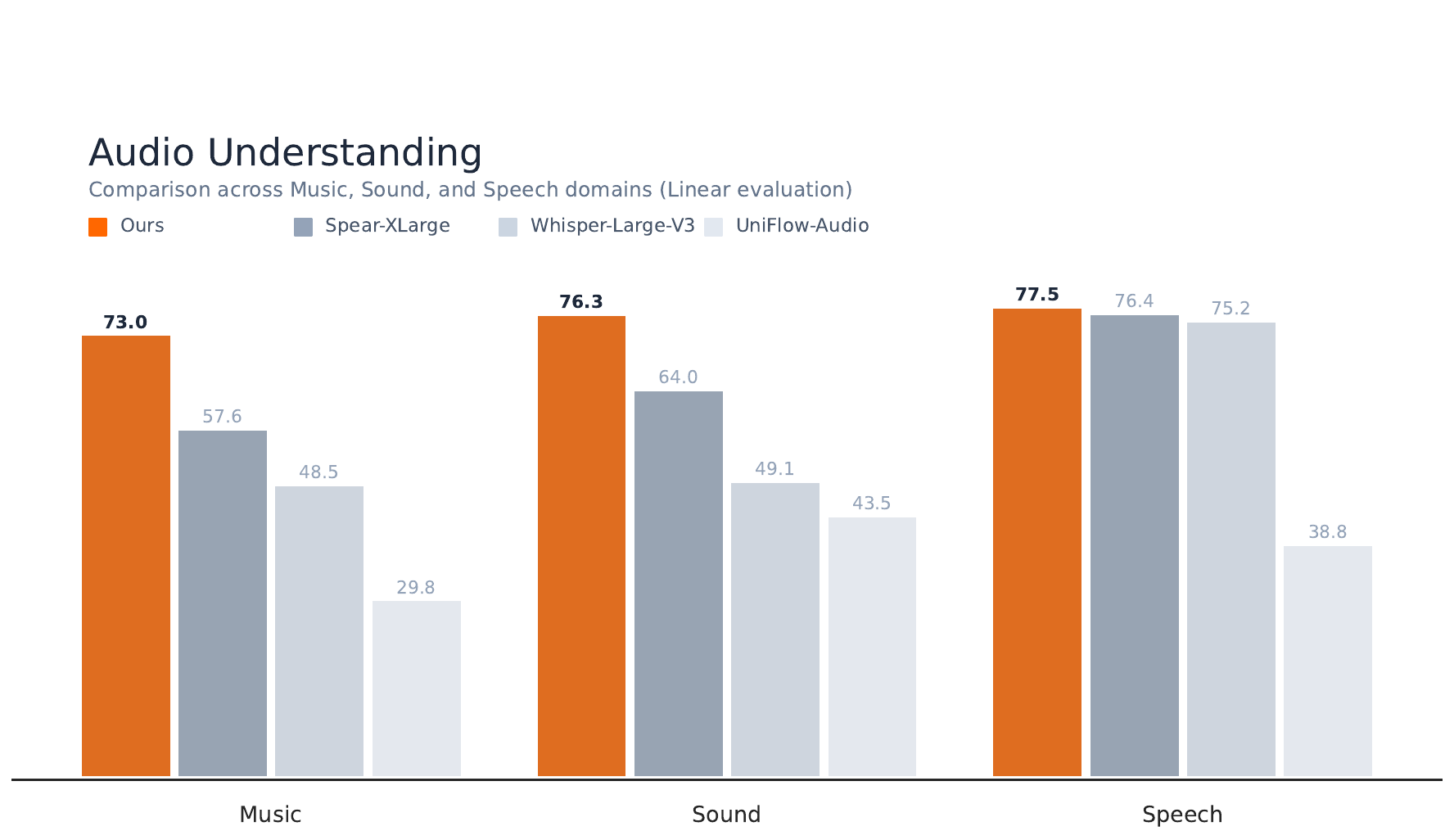}
         \caption{Audio Understanding Results}
         \label{fig:audio_results}
     \end{subfigure}
     \hfill 
     \begin{subfigure}[c]{0.49\linewidth}
         \centering
         \includegraphics[width=\linewidth]{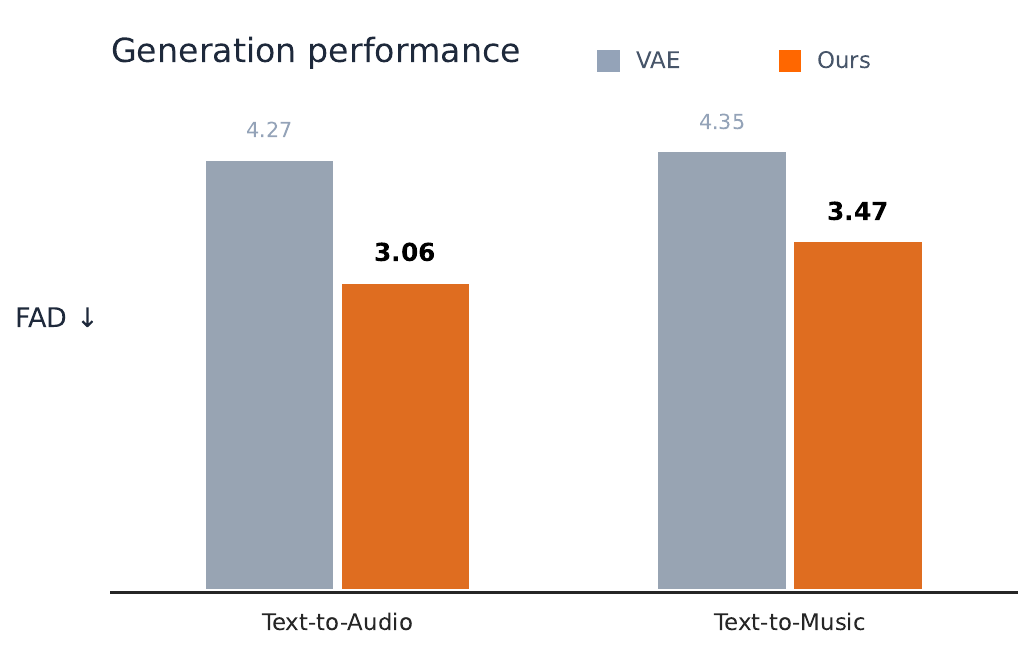}
         \caption{Generation Performance}
         \label{fig:second_graph}
     \end{subfigure}
     
     \caption{A summarization of DashengTokenizer capabilities for understanding and generation tasks.}
     \label{fig:combined_results}
\end{figure}

In contrast to previous single tokenizer methods, which distill high-dimensional semantic knowledge into a low-dimensional acoustic model, our approach aims to do the inverse: We embed low-dimensional acoustic information into high-dimensional semantic features.
We introduce DashengTokenizer, a unified continuous audio tokenizer designed for both understanding and generation across speech, music, and environmental sound domains.
Our approach freezes previous semantic knowledge from a pretrained, strong semantic encoder, and injects acoustic information from a mel-spectrogram via a linear projection. 
The method is simple, requiring to only train a the linear projection with an additional standard acoustic decoder.
A comparison with previous works can be seen in \Cref{tab:comparison_tokenizers}.
DashengTokenizer performs on par with previous continuous tokenizers for reconstruction tasks, while significantly outperforming codecs and encoders in general audio understanding.
Furthermore, our experiments in Text-to-Audio (TTA), Text-to-Music (TTM), and Speech Enhancement (SE) demonstrate the superior versatility of our representation for high-fidelity audio generation.
A summarization of DashengTokenizer's performance is seen in \Cref{fig:combined_results}.


\begin{table}[ht]
\centering
\caption{A comparison of our work in contrast to previous audio encoders and tokenizers.}
\label{tab:comparison_tokenizers}
\small
\resizebox{\linewidth}{!}{%
\begin{tabular}{r|cccccc}
\toprule
Approach & Type & Dimension & Representative work &   Understanding & Generation &  \\
\midrule
Audio Codec & Discrete & Low & VQ-VAE~\cite{van2017neural} & \xmark & \smark \\
Audio Codec + Semantic & Discrete & Low &  Mimi~\cite{kyutai2024moshi} & \smark & \smark \\
Audio Encoders & Continuous & High & Whisper~\cite{radford2023robust} & \cmark & \xmark  \\
Acoustic Tokenizers & Continuous & Low & AudioLDM~\cite{audioldm_haohe}  &  \xmark & \cmark \\
\midrule
Ours & Continuous & High & Ours & \cmark & \cmark \\
\bottomrule
\end{tabular}
}
\end{table}

\section{Previous works}

Current audio tokenization research primarily utilizes (Vector Quantized) Variational Autoencoders (VQ-VAE) to compress raw waveforms or spectrograms into low-dimensional latent spaces. 
While standard VAEs compress audio into a low-dimensional continuous latent, VQ-VAEs apply further quantization to produce discrete tokens suitable for sequence modeling, also known as codecs.

\paragraph{Codecs} A substantial body of literature concentrates on the compression of audio signals into discrete quantized tokens. 
Notably, foundational frameworks such as SoundStorm~\cite{zeghidour2021soundstream}, Encodec~\cite{defossez2022high} and DAC~\cite{dac} primarily prioritized maintaining high acoustic fidelity throughout the reconstruction process. 
However, with the rise of Large Language Models (LLMs) for speech and audio modeling~\cite{chu2024qwen2,dinkel2025midashenglm}, research has increasingly transitioned towards augmenting acoustic codecs with semantic information~\cite{kyutai2024moshi,chen2025sac,yangalmtokenizer,liu2024semanticodec,ye2025llasa,li2025dualcodec,zhang2024speechtokenizer,li2025flexicodec}.
This integration aims to provide the underlying LLM with richer contextual cues, enhancing performance in downstream understanding tasks.
While effective, due to the lossy quantization process, semantic codecs are outperformed in both understanding and reconstruction performance by unified tokenizers.

\paragraph{Unified tokenizers}

Developing a single representation capable of both high-level understanding and high-fidelity generation remains an open challenge.
While improvements have been made in recent years for general audio encoders~\cite{wang25m_interspeech_usam,yang2025spear,dinkel24b_interspeech,bharadwaj2025openbeats}, little emphasis has been paid on making these models capable of audio generation.

The work most closely related to ours is Ming-UniAudio~\cite{yan2025ming}.
However, DashengTokenizer differs in two fundamental aspects: (I) Domain Versatility: Ming-UniAudio is restricted to the speech domain, whereas our approach generalizes across speech, music, and environmental audio. (II) Simplicity: Ming-UniAudio requires a complex three-stage training pipeline (acoustic modeling, semantic distillation, and fine-tuning), mirroring the established semantic codec training pipeline.
In contrast, our framework employs a single-stage acoustic injection via a linear projection, significantly reducing training complexity while maintaining competitive performance.

\section{Methodology}
\label{sec:method}

\begin{figure}
    \centering
    \includegraphics[width=0.99\linewidth]{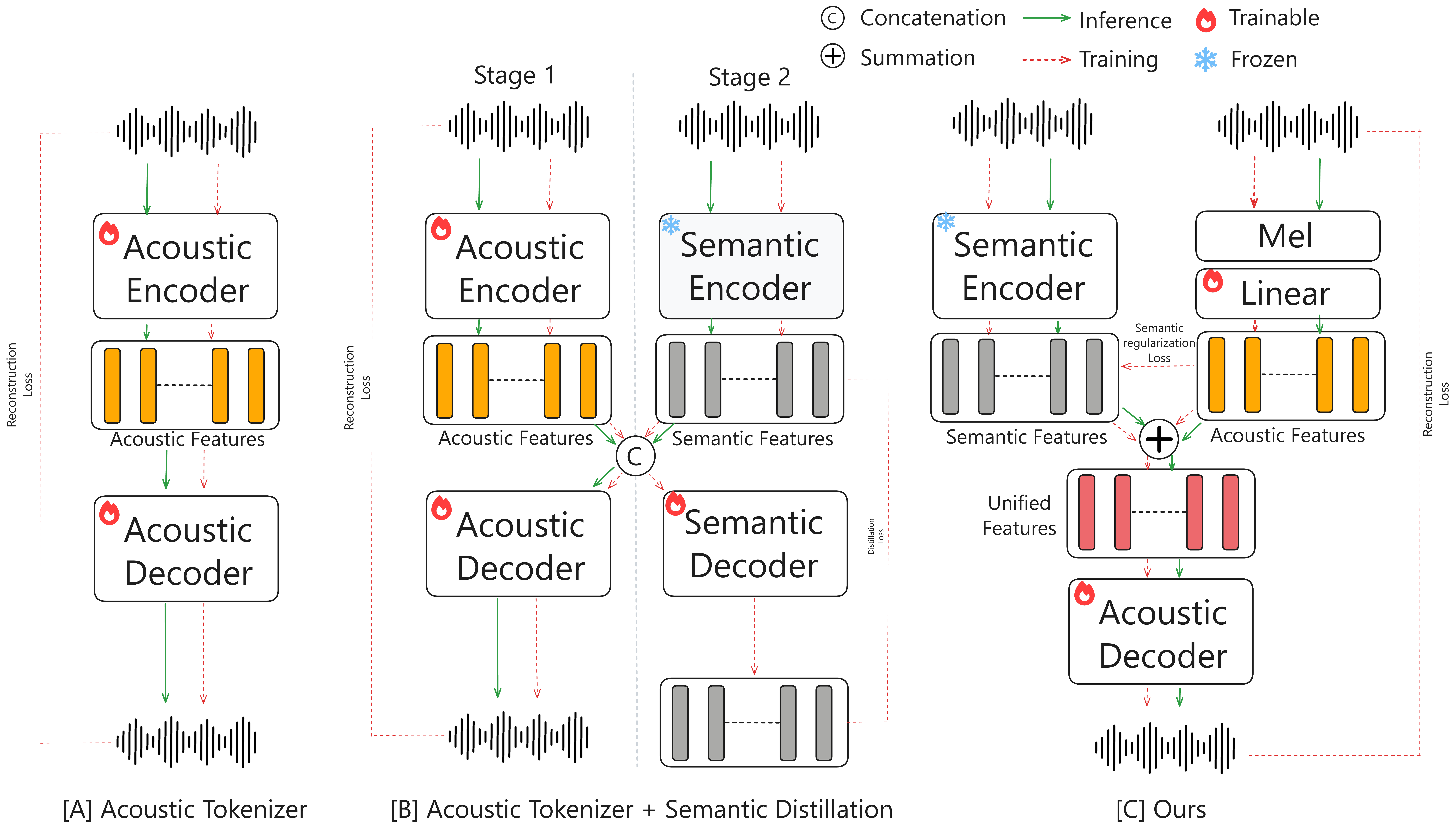}
    \caption{The proposed DashengTokenizer compared to prior approaches: [A] standard acoustic Acoustic modeling using VAE, and [B] semantically distilled (VQ-)VAEs. 
    In contrast, our approach eliminates the multi-stage training required by [B] and does not rely on a semantic decoder that is discarded during inference, thereby avoiding a train-test mismatch.}
    \label{fig:framework}
\end{figure}

Dashengtokenizer injects acoustic information into rich semantic embeddings.
A simplified overview between our approach and previous methods can be seen in \Cref{fig:framework}.

Given an input signal $x \in \mathbb{R}^{s}$, we first obtain a semantic features $z_{\text{sem}} \in \mathbb{R}^{b\times t\times d}$, where $b$, $t$, and $d$ represent batch size, temporal length, and feature dimension respectively, extracted from a frozen pre-trained model ($\mathcal{T}_{\text{frozen}}$) as:
\begin{equation}
\label{eq:semantic_extraction}
z_{\text{sem}} = \mathcal{T}_{\text{frozen}}(x).    
\end{equation}

Simultaneously, we extract acoustic embeddings $z_{\text{ac}} \mathbb{R}^{b\times t\times d}$ from the same input $x$, extracted via a combination of a MelSpectrogram, followed by a linear projection $\phi$ and layer normalization:
\begin{equation}
\label{eq:acoustic_extract}
z_{\text{ac}} = \text{LayerNorm}(\phi(\text{MelSpec}(x))).
\end{equation}

To ensure temporal alignment, $\phi$ is implemented as a non-overlapping patch embedding that maps consecutive spectrogram frames to the framerate of $z_{\text{sem}}$.
The final unified features $z$ is computed via additive fusion:
\begin{equation}
z = z_{\text{sem}} + z_{\text{ac}}.
\end{equation}

We train a generator (vocoder) $G$ to reconstruct the original input $x$ from $z$, $G(z) \mapsto x$.
The training follows a Generative Adversarial Network (GAN) framework using a Multi-Frequency Discriminator (MFD) \cite{ye2025llasa} and a hinge loss objective.
The generator loss $\mathcal{L}_G$ is defined as:
\begin{equation}
\label{eq:generator}
     \mathcal{L}_G = \lambda_{\text{sem}} \mathcal{L}_{\text{sem}} + \lambda_{\text{mel}} \mathcal{L}_{\text{mel}} + \mathcal{L}_{\text{fm}} + \mathcal{L}_{\text{adv}},
\end{equation}
where $\mathcal{L}_{\text{fm}}$ and $\mathcal{L}_{\text{adv}}$ represent feature-matching and adversarial losses, and $\mathcal{L}_{\text{mel}}$ is the $\ell_1$ reconstruction loss in the mel-frequency domain.
Crucially, we introduce a semantic preservation loss $\mathcal{L}_{\text{sem}}$ to prevent the acoustic features from overwhelming the semantic features, which would lead to a collapse of the understanding capabilities:
\begin{equation}
\mathcal{L}_{\text{sem}} = \| z_{\text{sem}} - z_{\text{ac}} \|_2^2.
\end{equation}
In our experiments, we set $\lambda_{\text{sem}} = 45$ and $\lambda_{\text{mel}} = 45$.
For more information we provide an ablation study in \Cref{sec:ablation_weight}.

As illustrated in \Cref{fig:framework}, DashengTokenizer offers superior efficiency through a single-stage training pipeline that eliminates the need for multi-stage distillation (see [B]) while ensuring inference consistency by retaining the semantic encoder during deployment.
Other benefits of our approach is that the acoustic feature is independent of the semantic feature, allowing generation performance to be enhanced via higher sampling rates or higher-resolution spectrograms as inputs, without retraining the semantic encoder.

\section{Experiments}
\label{sec:experiments}

\subsection{Datasets}
\label{ssec:train_dataset}

Our training pipeline utilizes approximately 282k hours of diverse audio data, sampled with specific domain weights: Music (21\%; Million Sound Dataset~\cite{bertin2011million}, MTG-Jamendo~\cite{bogdanov2019mtg}), English Speech (21\%; Emilia~\cite{he2025emilia}, Yodas~\cite{li2023yodas}, LibriLight~\cite{kahn2020libri}, CommonVoice15~\cite{commonvoice:2020}), Chinese Speech (40\%; AISHELL-1/2/3~\cite{bu2017aishell,du2018aishell2,shi2020aishell3}, Emilia), Other Languages (10\%; multi-lingual Emilia~\cite{he2025emilia}), and General Sound (26\%; AudioSet~\cite{gemmeke2017audioset}, FSD50K~\cite{fonseca2020fsd50k}, AudioCaps, CochlScene~\cite{jeong2022cochlscene}, ACAVCaps~\cite{niu2026acavcapsenablinglargescaletraining}).
By default, all datasets are resampled to 16 kHz.



\paragraph{Evaluation benchmarks}

We evaluate representations of DashengTokenizer across three distinct axes.
First, audio understanding is evaluated via the X-ARES~\cite{zhang2025xares} benchmark, utilizing a linear probing protocol across 22 tasks in the speech, music, and sound domains.
Second, reconstruction fidelity is quantified using wideband Perceptual Evaluation of Speech Quality (PESQ)~\cite{practicalpseq} and Short-Time Objective Intelligibility (STOI)~\cite{stoi} for speech on the SEED-TTS dataset~\cite{anastassiou2024seed}.
Signal reconstruction quality for music and environmental audio is measured via Mel-spectrogram (Mel-16k) and Short-Time Fourier Transform (STFT-16k) distances calculated via the DAC toolkit~\cite{dac} on MUSDB18-HQ~\cite{musdb18-hq} and AudioSet~\cite{gemmeke2017audioset}, respectively.
Finally, we validate generative capabilities through downstream experiments in Speech Enhancement (Valentini~\cite{ValentiniBotinhao2017NoisySD}, DNS~\cite{dns1}), Text-to-Audio (AudioCaps~\cite{kim2019audiocaps}), and Text-to-Music (MusicCaps~\cite{agostinelli2023musiclm}).
For SE, we employ DNSMOS P.835~\cite{reddy2022dnsmos}, {NISQAv2}~\cite{mittag21_interspeech}, and \text{Spksim} to assess signal quality and speaker identity preservation. 
For both TTA and TTM, we evaluate performance using Fr\'{e}chet Audio Distance (\text{FAD}~\cite{kilgour19_interspeech}), Fr\'{e}chet Distance (\text{FD}), Kullback–Leibler (\text{KL}) divergence, and \text{CLAPScore}~\cite{wu2023large} to measure text-audio alignment.
\subsection{Setup}
\label{ssec:setup}

The model is trained for one million steps using the AdamW~\cite{loshchilovdecoupled} optimizer with a global batch size of 256. 
The learning rate is initialized at $5 \times 10^{-4}$ and decayed to 10\% using a cosine schedule.

\subsubsection{Architecture Configuration}

Our framework consists of a semantic encoder, a trained acoustic encoder and an acoustic decoder, summarized in~\Cref{tab:model_specs}. We further discuss each component individually.

\begin{table}[t]
\centering
\small
\caption{DashengTokenizer architecture.}
\label{tab:model_specs}
\begin{tabular}{r|lll}
\toprule
{Attribute} & {Semantic Encoder} & {Acoustic Encoder} & {Decoder} \\ 
\midrule
Parameters & 630 M & 0.66 M & 173 M \\
Architecture & 32 Layer Transformer & 2D Conv & Vocos \\
Hidden Dim & 1280 & 1280 & 1280 \\
Input & 64-bin Mel & 128-bin Mel & Unified ($d=1280$) \\
Frame Rate & 25 Hz & 25 Hz & 25 Hz (Upsample to 50 Hz) \\
\bottomrule
\end{tabular}
\end{table}

\paragraph{Semantic encoder}

We utilize the 630M parameter encoder from MiDashengLM-7B~\cite{dinkel2025midashenglm}, a 32-layer Transformer pretrained for general audio understanding on public datasets.
Unlike alternatives with unknown training data mixture like Whisper~\cite{radford2023robust}, the MiDashengLM backbone ensures full reproducibility. It operates at a 25 Hz frame rate with $d=1280$ dimensional embeddings.

\paragraph{Acoustic Encoder}

To inject acoustic information, we apply a 2D convolution over a 128-bin mel-spectrogram, extracted every 10 ms. 
We use non-overlapping patches ($128 \times 4$ bins) to match the 25 Hz temporal resolution of the semantic feature. 
This lightweight component (0.66M parameters) followed by LayerNorm~\cite{ba2016layer} produces the residual features $z_{\text{ac}}$.

\paragraph{Acoustic decoder}

The decoder upsamples the unified 25 Hz features to 50 Hz via a 1D transposed convolution. 
We adopt a scaled Vocos architecture \cite{siuzdak2023vocos} (173M parameters, 12 layers, 1280 hidden dimension) to accommodate the high-dimensional latent space ($d=1280$).

\section{Results}
\label{sec:results}

We evaluate our results against existing literature, which can be broadly categorized into three distinct approaches: neural discrete codecs, audio encoders, and continuous tokenizers.

\subsection{Reconstruction evaluation}

We evaluate speech reconstruction quality on the Mandarin (ZH) and English (EN) subsets of the Seed-TTS benchmark~\cite{anastassiou2024seed}.
Here, we compare against speech codecs and continuous acoustic tokenizers (EzAudio~\cite{hai2024ezaudio}, UniFlow-Audio~\cite{xu2025uniflow}).
All waveform targets and model outputs are resampled to 16 kHz for consistent evaluation.

The results, summarized in \Cref{tab:reconstuction_speech}, demonstrate that continuous tokenizers generally exceed the performance of discrete codecs across both language benchmarks. 
While high-performance codecs like DAC~\cite{dac} achieve strong results, they are consistently outperformed by continuous representations in perceptual quality.
Notably, our proposed model achieves competitive results, while maintaining a lower framerate (25 Hz) compared to other top-performing continuous models.

\begin{table}[t]
\centering
\caption{Speech reconstruction performance across different tokenizers. Codecs are highlighted in \colorbox{xiaomiblue!5}{blue}, while continuous tokenizers are highlighted in \colorbox{red!5}{red}. For all metrics higher is better and best are in \textbf{bold} and second best are \underline{underlined}.}
\label{tab:reconstuction_speech}
\begin{tabular}{rr|cccc}
\toprule
Model &  Framerate & \multicolumn{2}{c}{Seed-TTS (ZH)} & \multicolumn{2}{c}{Seed-TTS (EN)} \\
\cmidrule(lr){3-4} \cmidrule(lr){5-6}
& & PESQ $\uparrow$ & STOI $\uparrow$ & PESQ $\uparrow$ & STOI $\uparrow$ \\
\midrule

 \rowcolor{xiaomiblue!5}  SNAC~\cite{siuzdak2024snac}                      & - & 1.841 & 0.862 & 1.804 & 0.870 \\
 \rowcolor{xiaomiblue!5}  Mimi~\cite{kyutai2024moshi}                      & 12.5 & 2.050 & 0.890 & 2.010 & 0.890 \\
 \rowcolor{xiaomiblue!5}  XCodec 2.0~\cite{ye2025llasa}                & 50 & 2.190 & 0.920 & 2.370 & 0.930 \\
 \rowcolor{xiaomiblue!5}  XY-Tokenizer~\cite{gong2025xy}              & 12.5 & 2.270 & 0.900 & 2.140 & 0.900 \\
 \rowcolor{xiaomiblue!5}  DAC~\cite{dac}                      & 50  & 3.860 & 0.967 & 3.763 & 0.969 \\
 
\midrule
\rowcolor{xiaomired!5}  EzAudio~\cite{hai2024ezaudio}           & 50 & 3.857 & 0.987  & 3.668 &	\underline{0.989} \\
\rowcolor{xiaomired!5}  UniFlow-Audio~\cite{xu2025uniflow}           & 50 & 4.048 & \textbf{0.990}  & 3.858  &	\textbf{0.992} \\
\rowcolor{xiaomired!5}  MingTok-Audio~\cite{yan2025ming}             & 50 & \textbf{4.210} & 0.980 & \underline{4.040} & 0.980 \\
\rowcolor{xiaomired!5} Ours                      & 25 & \underline{4.163} & \underline{0.988} & \textbf{4.125}& {0.987} \\
\bottomrule
\end{tabular}
\end{table}

Beyond speech tasks, \Cref{tab:reconstruction_general_audio} evaluates reconstruction performance on general environmental audio and music. 
Notably, our model achieves strong results for AudioSet in the Mel-16k metric with a score of 0.320, outperforming baselines. 
While UniFlow-Audio-VAE remains highly competitive in music reconstruction, our model demonstrates robust generalization across diverse acoustic environments, consistently placing among the top-tier performers.

\begin{table}[htb]
\centering
\caption{General audio (Audioset) and music (MUSDB) reconstruction performance measured with Mel and STFT distances on 16 kHz audio. Codecs are highlighted in \colorbox{xiaomiblue!5}{blue}, while continuous tokenizers are highlighted in \colorbox{xiaomired!5}{red}. For all metrics lower is better and best are in \textbf{bold} and second best are \underline{underlined}.}
\label{tab:reconstruction_general_audio}
\begin{tabular}{rr|cccc}
\toprule
Model & Framerate &  \multicolumn{2}{c}{Audioset} & \multicolumn{2}{c}{MUSDB18} \\
\cmidrule(lr){3-4} \cmidrule(lr){5-6}
& & Mel-16k $\downarrow$ & STFT-16k  $\downarrow$ & Mel-16k $\downarrow$ & STFT-16k  $\downarrow$ \\
\midrule
 \rowcolor{xiaomiblue!5} XY-Tokenizer & 12.5    & 0.901 & 2.361 & 0.607 & 1.817 \\
\rowcolor{xiaomiblue!5}  DAC & 50             & 0.664 & 1.986 & 0.965 & 2.352 \\
\rowcolor{xiaomiblue!5}  SNAC & -            & 1.195 & 2.690 & 1.242 & 2.655 \\
\rowcolor{xiaomiblue!5}  XCodec 2.0 &  50    & 1.253 & 2.925 & 1.273 & 2.855 \\
\midrule
\rowcolor{xiaomired!5}  EzAudio & 50  & 0.349 & \underline{1.434} & \underline{0.258}  & \underline{1.168} \\
\rowcolor{xiaomired!5}  UniFlow-Audio & 50 & 0.321 & \textbf{1.385} & \textbf{0.232} & \textbf{1.129} \\

\rowcolor{xiaomired!5}  MingTok-Audio  & 50   & 0.462 & 1.768 & 0.432 & 1.616 \\
\rowcolor{xiaomired!5}  Ours     & 25       & \textbf{0.320} & {1.585} & {0.293} & {1.458} \\
\bottomrule
\end{tabular}
\end{table}

\subsection{Understanding}
\label{ssec:understanding}

\paragraph{Speech understanding}

Speech understanding performance on the X-ARES benchmark is reported across eleven tasks, being emotion recognition (Crema-D, RAV), intent classification (FSC), speaker identification (VoxC), speaker counting (LibCnt), keyword spotting (SPV1), automatic speech recognition (LS100h), gender classification (LSMF), vocal sound classification (VocS), vocal imitation classification (VocI) and language identification (VoxL33).
For all discrete codecs, features are extracted as continuous latents prior to the quantization stage.
Note that MingTok-Audio~\cite{yan2025ming} operates in two modes: A low dimensional Acoustic VAE embedding and a high-dimensional unified embedding, both of which are evaluated here.

\begin{table}[htb]
\centering
\caption{Understanding Performance in the Speech Domain on the X-ARES benchmark. Codecs are highlighted in \colorbox{xiaomiblue!5}{blue}, while encoder models are highlighed in \colorbox{yellow!5}{yellow} and continuous tokenizers are highlighted in \colorbox{xiaomired!5}{red}. For all metrics higher is better and best are in \textbf{bold} and second best are \underline{underlined}.}
\label{tab:speech_understanding}
\begin{adjustbox}{width=\textwidth}
\begin{tabular}{r|ccccccccccc}
\toprule
{Model} & {CREMA-D} & {FSC} & {LibCnt} & {LS100h} & {LSMF} & {RAV} & {SPV1} & {VocI} & {VocS} & {VoxC} & {VoxL33} \\ \midrule
\midrule
\rowcolor{xiaomiblue!5} DAC~\cite{dac}  & 43.90 & 2.55  & 43.93 & 0.00   & 91.58 & 36.45 & 15.50 & 2.23  & 42.17 & 13.24 & 7.96  \\
\rowcolor{xiaomiblue!5}  Encodec~\cite{defossez2022high}  & 39.16 & 3.48  & 33.78 & 0.00   & 88.44 & 27.71 & 26.70 & 2.16  & 43.06 & 10.92 & 7.89  \\
\rowcolor{xiaomiblue!5} SemantiCodec~\cite{liu2024semanticodec}  & 60.30 & 46.40 & 66.05 & 0.00   & 98.14 & 55.00 & 88.02 & 23.82 & 84.28 & 60.99 & 26.66 \\
\rowcolor{xiaomiblue!5}  FlexiCodec~\cite{li2025flexicodec} & 58.00 & \textbf{99.00} & 43.00 & 0.00   & 0.00  & 50.00 & 0.00  & 0.00  & 79.00 & 0.00  & 72.00 \\
\rowcolor{xiaomiblue!5}  WavTokenizer~\cite{jiwavtokenizer} & 33.00 & 2.00  & 38.00 & 0.00   & 0.00  & 27.00 & 0.00  & 0.00  & 27.00 & 0.00  & 9.00  \\
\midrule
\rowcolor{yellow!5} OpenBeats~\cite{bharadwaj2025openbeats} & 65.90 & 58.50 & 69.90 & 0.00   & 97.30 & 63.00 & 94.40 & 21.40 & 87.70 & 0.00  & 0.00  \\
\rowcolor{yellow!5} Whisper-Large-V3~\cite{radford2023robust} & 71.32 & 97.78 & 64.40 & \underline{90.00}  & 94.93 & 68.47 & \textbf{97.75} & 29.30 & 91.48 & 24.76 & \textbf{97.38} \\
\rowcolor{yellow!5} HuBERT~\cite{hsu2021hubert} & 58.06 & \underline{98.73} & 49.72 & 82.45  & 85.94 & 47.71 & 95.67 & 18.82 & 81.71 & 3.39  & 45.06 \\
\rowcolor{yellow!5} MSM-MAE~\cite{niizumi2022masked} & 68.64 & 0.47  & \underline{71.22} & 0.00   & 97.58 & 66.39 & 92.77 & 0.34  & 87.56 & 71.64 & 66.63 \\
\rowcolor{yellow!5} WavLM~\cite{chen2022wavlm} & 45.16 & 96.07 & 50.37 & 67.06  & 75.95 & 32.57 & 89.77 & 10.71 & 71.19 & 4.51  & 40.89 \\
\rowcolor{yellow!5} SPEAR-Base~\cite{yang2025spear} & 72.90 & 96.90 & 62.40 & \underline{90.00}  & \underline{98.60} & 68.80 & 97.10 & 29.10 & 90.20 & 65.50  & 86.10 \\
\rowcolor{yellow!5} SPEAR-XLarge~\cite{yang2025spear} & 78.41 & 98.20 & 69.00 & 88.43 & \textbf{98.73}&  \underline{76.59} & 97.30 &\textbf{31.28} & 92.62 & \underline{79.42} &  90.49   \\
\rowcolor{yellow!5} Dasheng-Base~\cite{dinkel24b_interspeech}  & 77.20 & 94.38 & 68.80 & 72.40  & 98.47 & 72.50 & 96.70 & 25.05 & 91.00 & 77.98 & 81.29 \\
\rowcolor{yellow!5} Dasheng 0.6B~\cite{dinkel24b_interspeech}  & \underline{79.39} & 97.79 & 69.67 & 69.64  & {98.58} & 78.13 & 97.22 & \underline{29.67} & \underline{92.21} & \textbf{84.00} & 86.33 \\
\midrule
\rowcolor{xiaomired!5} EzAudio~\cite{hai2024ezaudio} & 38.80 & 1.63 &  37.19 & 0.00 &  84.90 & 25.56 & 12.47 &  1.21 &  35.26 & 5.99 & 6.96 \\
\rowcolor{xiaomired!5} UniFlow-Audio~\cite{xu2025uniflow} & 39.78 & 1.32 & 37.15 & 0.00 &  74.57 & 23.26 & 11.70 &  1.37 & 34.73 & 6.18 & 6.34 \\

\rowcolor{xiaomired!5} MingTok-Audio (Acoustic)~\cite{yan2025ming}  & 41.22 & 2.82  & 39.42 & 0.00   & 86.91 & 30.83 & 20.34 & 2.45  & 38.63 & 6.97  & 9.88  \\
\rowcolor{xiaomired!5} MingTok-Audio (Unified)~\cite{yan2025ming}  & 66.85 & 98.58 & 62.92 & \textbf{93.45}  & 97.10 & 57.08 & 96.14 & 20.64 & 88.59 & 35.28 & 77.44 \\
\rowcolor{xiaomired!5} Ours  & \textbf{80.56} & 83.39 & \textbf{79.98} & 68.71  & 98.06 & \textbf{83.06} & \underline{97.35} & 23.92 & \textbf{93.15} & 51.16 & \underline{93.60} \\
\bottomrule
\end{tabular}
\end{adjustbox}
\end{table}

The results in \Cref{tab:speech_understanding} reveal a distinct representational gap where acoustic (VQ-)VAEs like EzAudio and DAC excel at reconstruction but lack the high-level features necessary for discriminative tasks.
Our proposed DashengTokenizer offers highly competitive performance, securing the best score on four tasks (CREMA-D, LibCnt, RAV, VocS) and ranks second on two others (SPV1, VoxL33). 
The notable performance gap of DashengTokenizer observed on FSC and LS100h likely stems from our model's retention of acoustic variance; while this is beneficial for emotion and scene recognition, it can interfere with tasks requiring pure semantic abstraction.
Crucially, unlike audio encoders such as SPEAR-XLarge, our unified framework maintains high-fidelity synthesis capabilities, effectively bridging the divide between audio understanding and generation.

\paragraph{Sound understanding}

We further evaluate the representational capacity of DashengTokenizer across diverse sound understanding tasks, including spoofing detection (ASV), sound-to-text retrieval (Clo), sound/environment classification (ESC, Urb8, FSD50, F18K), and domestic sound event detection (DES). 
The results in \Cref{tab:sound_understanding} mirror the trends observed in the speech domain (\Cref{tab:speech_understanding}), where neural codecs exhibit a significant performance plateau on semantic tasks.

In contrast, DashengTokenizer demonstrates performance leadership across the majority of the sound understanding benchmark. 
Our model achieves scores of $96.40$ on ESC, $59.98$ on FSD50, $86.81$ on F18K, and $55.40$ on DES. 
While our framework marginally trails SPEAR-XLarge in spoofing detection (ASV), it achieves substantial gains in sound-to-text retrieval (Clo) with a score of $5.64$. 

\begin{table}[htb]
\centering
\caption{Understanding Performance in the Sound Domain on the X-ARES benchmark. Codecs are highlighted in \colorbox{xiaomiblue!5}{blue}, while continuous tokenizers are highlighted in \colorbox{xiaomired!5}{red} and encoder models are highlighed in \colorbox{yellow!5}{yellow}. For all metrics higher is better and best are in \textbf{bold} and second best are \underline{underlined}.}
\label{tab:sound_understanding}
\begin{tabular}{r|ccccccc}
\toprule
{Model} & {ASV} & {Clo} & {DES} & {ESC} & {FSD50} & {F18K} & {Urb8} \\ \midrule
\rowcolor{xiaomiblue!5} DAC & 93.46 & 0.73  & 20.20 & 31.20 & 7.80  & 16.44 & 50.26 \\
\rowcolor{xiaomiblue!5} Encodec & 95.14 & 0.57  & 3.82  & 16.80 & 4.63  & 6.94  & 40.34 \\
\rowcolor{xiaomiblue!5} SemantiCodec & 93.69 & 1.89  & 48.57 & 81.75 & 34.84 & 26.69 & 84.54 \\
\rowcolor{xiaomiblue!5} FlexiCodec & 93.00 & 0.00  & 17.00 & 30.00 & 0.00  & 0.00  & 0.00  \\
\rowcolor{xiaomiblue!5} WavTokenizer & 95.00 & 0.00  & 16.00 & 19.00 & 0.00  & 0.00  & 0.00  \\
\midrule
\rowcolor{yellow!5} OpenBeats & 0.00  & \underline{4.00}  & 55.20 & 86.80 & 38.00 & \underline{68.90} & \textbf{86.30} \\
\rowcolor{yellow!5} Whisper-Large-V3 & 97.94 & 3.10  & 22.64 & 62.45 & 32.00 & 49.56 & 75.74 \\
\rowcolor{yellow!5} HuBERT & 96.24 & 1.74  & 3.64  & 37.35 & 17.40 & 26.50 & 57.40 \\
\rowcolor{yellow!5} MSM-MAE & 93.19 & 3.86  & 0.00  & \underline{88.70} & 44.15 & 48.63 & 84.35 \\
\rowcolor{yellow!5} WavLM & 95.37 & 0.79  & 16.76 & 31.45 & 10.06 & 19.63 & 53.13 \\
\rowcolor{yellow!5} SPEAR-Base & 99.20 & 3.65  & 48.63 & 82.25 & 37.96 & 48.00 & 81.46 \\
\rowcolor{yellow!5} SPEAR-XLarge & \textbf{99.81} & 3.37  &  53.42 & {88.45} & 44.55 & 64.82 & 83.78 \\
\rowcolor{yellow!5} Dasheng-Base & 96.29 & 3.30  & 53.20 & 86.90 & 40.90 & 55.70 & 83.50 \\
\rowcolor{yellow!5} Dasheng 0.6B & \underline{99.24} & \underline{4.00}  & \underline{55.01} & 87.95 & \underline{44.79} & 61.63 & \underline{84.71} \\
\midrule
\rowcolor{xiaomired!5} EzAudio & 91.44 & 0.40 & 17.01 & 18.05 & 4.86 & 13.38 & 34.97  \\
\rowcolor{xiaomired!5} UniFlow-Audio  & 93.77 &  0.57 & 16.90 & 19.85 & 4.83 & 12.19 &  35.16\\
\rowcolor{xiaomired!5} MingTok-Audio (Acoustic) & 92.41 & 0.63  & 15.98 & 23.15 & 5.82  & 13.88 & 44.85 \\
\rowcolor{xiaomired!5} MingTok-Audio (Unified) & 98.70 & 2.00  & 39.80 & 62.25 & 24.83 & 40.81 & 72.87 \\
\rowcolor{xiaomired!5}{Ours} & {96.45} & \textbf{5.64} & \textbf{55.40} & \textbf{96.40} & \textbf{59.98} & \textbf{86.81} & {83.80} \\
\bottomrule
\end{tabular}
\end{table}

\paragraph{Music understanding}

The results for music understanding, presented in \Cref{tab:music_understanding}, continue the trends observed in the speech and general sound domains.
Encoder models generally set the strongest baseline, while neural audio codecs perform poorly. 
Our proposed DashengTokenizer outperforms all others approaches on three of four datasets, with a particularly large lead on the challenging MAESTRO benchmark. 
This result exemplifies that acoustic information present in pretrained embeddings, can be helpful in the music domain.
Additional experimentation in regards to the effectiveness of acoustic injection can be seen in \Cref{tab:understanding_ablation}.

\begin{table}[htb]
\centering
\caption{Understanding Performance in the Music Domain on the X-ARES benchmark. Codecs are highlighted in \colorbox{xiaomiblue!5}{blue}, while continuous tokenizers are highlighted in \colorbox{xiaomired!5}{red} and encoder models are highlighed in \colorbox{yellow!5}{yellow}. For all metrics higher is better and best are in \textbf{bold} and second best are \underline{underlined}.}
\label{tab:music_understanding}
\begin{tabular}{r|cccc}
\toprule
{Model} & {FMA} & {GTZAN} & {MAESTRO} & {NSynth} \\
\midrule
\rowcolor{xiaomiblue!5} DAC & 35.43 & 54.06 & 12.78 & 38.57 \\
\rowcolor{xiaomiblue!5} Encodec & 28.80 & 35.14 & 23.37 & 32.59 \\
\rowcolor{xiaomiblue!5} SemantiCodec & 57.94 & 66.67 & 8.62 & 68.14 \\
\rowcolor{xiaomiblue!5} FlexiCodec & 0.00 & 0.00 & 0.00 & 0.00 \\
\rowcolor{xiaomiblue!5} WavTokenizer & 0.00 & 0.00 & 0.00 & 0.00 \\
\midrule
\rowcolor{yellow!5} OpenBeats & 62.40 & 85.90 & 0.00 & 58.90 \\
\rowcolor{yellow!5} Whisper-Large-V3 & 58.90 & 71.77 & 0.00 & 63.50 \\
\rowcolor{yellow!5} HuBERT & 42.81 & 51.85 & 0.00 & 48.22 \\
\rowcolor{yellow!5} MSM-MAE & 62.74 & 84.39 & 0.32 & 73.02 \\
\rowcolor{yellow!5} WavLM & 36.11 & 48.14 & 0.00 & 40.14 \\
\rowcolor{yellow!5} SPEAR-Base & 62.06 & 81.08 & 12.65  &  63.55 \\
\rowcolor{yellow!5} SPEAR-XLarge & 64.16 & 84.59 &  13.63 & 67.85 \\
\rowcolor{yellow!5} Dasheng-Base & 64.00 & 86.90 & 43.25 & 69.29 \\
\rowcolor{yellow!5} Dasheng 0.6B & \underline{64.34} & \underline{89.49} & \underline{46.81} & \textbf{78.39} \\
\midrule
\rowcolor{xiaomired!5} EzAudio & 30.74 & 40.54 & 17.57 & 32.18 \\
\rowcolor{xiaomired!5} UniFlow-Audio & 30.17 & 40.74 & 16.44 & 31.73 \\
\rowcolor{xiaomired!5} MingTok-Audio (Acoustic) & 34.74 & 39.94 & 21.38 & 27.81 \\
\rowcolor{xiaomired!5} MingTok-Audio (Unified) & 54.45 & 71.17 & 25.00 & 57.00 \\
\rowcolor{xiaomired!5} {Ours} & \textbf{66.29} & \textbf{89.99} & \textbf{57.65} & \underline{77.91} \\
\bottomrule
\end{tabular}
\end{table}

\subsection{Speech enhancement}

We evaluate the utility of DashengTokenizer for speech enhancement (SE) by employing the latent-space denoising framework described in \cite{sun25g_interspeech}.
We generate noisy speech by mixing clean samples with additive noise, then convert both into embeddings using our tokenizer. 
A lightweight 3-layer Transformer denoiser is trained to map these noisy embeddings back to the clean manifold, optimizing for the mean squared error between the enhanced and original clean embeddings.
To ensure a fair comparison, our experimental protocol, our training and evaluation setup is identical to the one in \cite{sun25g_interspeech}.

\begin{table}[htb]
\centering
\caption{Speech enhancement results on the Valentini and DNS1 datasets. Baselines are taken from~\cite{sun25g_interspeech}. Higher is better and best is in \textbf{bold}.}
\label{tab:speech_enhancement}

\resizebox{\linewidth}{!}{%
\begin{tabular}{llccccccccc}

\toprule
\multirow{2}{*}{{Dataset}} & \multirow{2}{*}{{Model}} & \multirow{2}{*}{{PESQ} $\uparrow$} & \multirow{2}{*}{{STOI} $\uparrow$} & \multicolumn{4}{c}{{DNSMOS} $\uparrow$} & \multirow{2}{*}{{NISQAv2} $\uparrow$} & \multirow{2}{*}{SpkSim $\uparrow$}  \\ 
\cmidrule(lr){5-8}
 &  &  &  & {P808} & {SIG} & {BAK} & {OVL} &  \\ 
\midrule
\rowcolor{xiaomilightgray!10} \multirow{6}{*}{Valentini} &  Noisy & 1.97 & 0.92 & 3.09 & 3.32 & 3.11 & 2.68 & 3.04 &  
{0.888} \\
 & LMS & 1.77 & 0.85 & 3.45 & 3.12 & 4.07 & 2.90 & 4.25 & 0.718 \\
 & Whisper  & 1.31 & 0.80 & 3.39 & 3.47 & 4.12 & 3.22 & 3.83 & 0.406 \\
 & WavLM  & 1.27 & 0.81 & 3.49 & \textbf{3.49} & 4.13 & \textbf{3.26} & 4.12 & 0.408 \\
 & Dasheng Denoiser & 2.32 & 0.90 & \textbf{3.52} & 3.45 & \textbf{4.14} & 3.22 & 4.51 & 0.783 \\ 
 \cmidrule{3-10}
 & {Ours} & \textbf{2.62} & \textbf{0.92} & 3.50 & 3.33 & 4.08 & 3.09 & \textbf{4.58} & \textbf{0.814} \\ 
\midrule
\rowcolor{xiaomilightgray!10} \multirow{6}{*}{DNS1} & Noisy & 1.58 & 0.92 & 3.16 & 3.39 & 2.62 & 2.48 & 2.58 & 0.924 \\
 & LMS & 1.90 & 0.90 & 3.87 & 3.34 & 4.07 & 3.10 & 3.85 & 0.847 \\
 & Whisper  & 1.20 & 0.81 & 3.87 & 3.60 & 4.14 & 3.36 & 4.09 & 0.489 \\
 & WavLM  & 1.20 & 0.83 & 3.91 & \textbf{3.61} & \textbf{4.18} & \textbf{3.40} & 4.41 & 0.486 \\
 & Dasheng Denoiser & 2.24 & 0.92 & \textbf{4.03} & 3.58 & 4.17 & 3.37 & 4.41 & {0.881} \\ 
 \cmidrule{3-10}
  & {Ours} & \textbf{2.66} & \textbf{0.95} & 4.02 & 3.52 & 4.16 & 3.30 & \textbf{4.46} & \textbf{0.902} \\ \bottomrule
\end{tabular}
}

\end{table}

Results in \Cref{tab:speech_enhancement} resents a comparative evaluation of our proposed DashengTokenizer against several baselines, including a log-Mel spectrogram (LMS) baseline, common speech encoder models (Whisper, WavLM), and Dasheng Denoiser~\cite{sun25g_interspeech}. 
Our model consistently outperforms all baselines in objective speech quality and intelligibility metrics. 
Specifically, on both Valentini and DNS1 datasets, DashengTokenizer significantly exceeds the performance of Dasheng Denoiser in terms of PESQ and STOI.
Although certain baselines show marginal leads in specific DNSMOS sub-metrics, our model yields the highest NISQAv2 scores (4.58 and 4.46) and the highest speaker similarity across both datasets, indicating superior overall perceptual quality.

\subsection{Flow based audio generation}

\begin{table}[htb]
\caption{Flow-matching generation performance for TTA and TTM, compared with the VAE from UniFlow-Audio. Best results are marked in \textbf{bold}.}
\label{tab:flow_results}
\centering
\begin{tabular}{rrcccc} 
\toprule
Task & Model & FAD $\downarrow$ & FD $\downarrow$ & KL $\downarrow$ & CLAP Score $\uparrow$ \\ 
\midrule
\multirow{2}{*}{TTA} &   VAE  & 4.27          & 24.72          & 1.84          & 0.412          \\ 
& Ours & \textbf{3.06} & \textbf{21.23} & \textbf{1.56} & \textbf{0.438} \\
\midrule
\multirow{2}{*}{TTM} &   VAE  & 4.35          & 23.68          & \textbf{1.94} & 0.258          \\
& Ours & \textbf{3.47} & \textbf{22.74} & 2.00          & \textbf{0.261} \\
\bottomrule
\end{tabular}
\end{table}

We evaluate the performance of our proposed framework on text-to-music (TTM) and text-to-audio (TTA) generation tasks. 
Our architecture adopts the flow-based DiT~\cite{peebles2023scalable} paradigm established by UniFlow-Audio~\cite{xu2025uniflow}, by replacing the baseline Variational Autoencoder (VAE) with our DashengTokenizer.

Because our unified latent representation resides in a significantly higher-dimensional space ($d=1280$) compared to the standard VAE ($d=128$), we follow RAE~\cite{zheng2025diffusion} and scale the DiT width from 1024 to 1532.
To keep the model capacity comparable to the baseline, we simultaneously reduce the DiT layers from 24 to 11, resulting in approximately 750M trainable parameters for both models.

For both TTM and TTA tasks, models are trained for 200k iterations using a global batch size of 192 distributed across eight NVIDIA GPUs. 
We utilize LP-MusicCaps~\cite{doh2023lp} and WavCaps~\cite{mei2024wavcaps} as the primary training corpora for the music and audio domains and evaluate on MusicCaps~\cite{agostinelli2023musiclm} and AudioCaps~\cite{kim2019audiocaps}, respectively. 
During inference, we use 25 inference steps with a classifier-free guidance (CFG) scale of 5.0.

As presented in \cref{tab:flow_results}, our proposed method demonstrates superior generation capabilities compared to the VAE baseline across both tasks.

\begin{figure}[htb]
    \centering
    \includegraphics[width=1.01\linewidth]{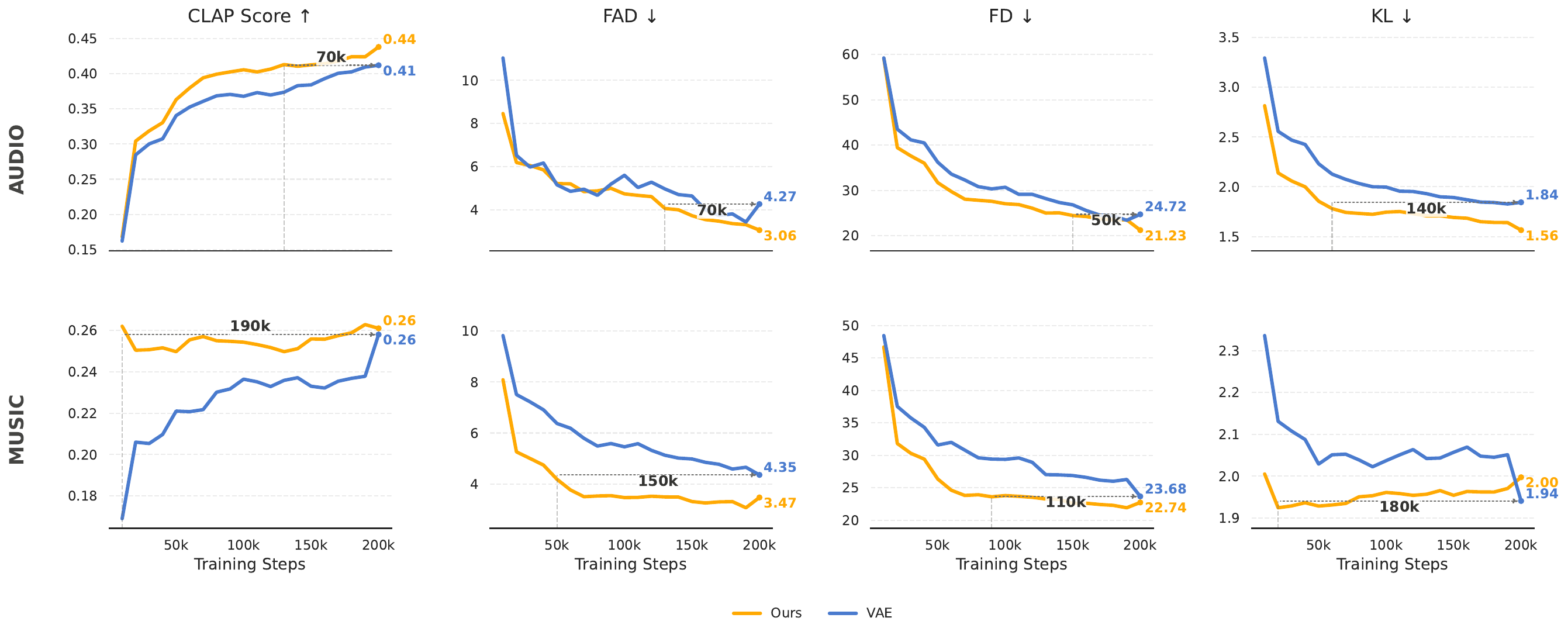}
    \caption{Text-to-Audio and Text-to-Music training progress of our proposed framework compared with the VAE from UniFlow-Audio. }
    \label{fig:evaluation_grid}
\end{figure}

Furthermore, the training progress illustrated in \Cref{fig:evaluation_grid} demonstrates that our proposed DashengTokenizer converges significantly faster than the baseline VAE. 
Notably, DashengTokenizer reaches comparable performance to the baseline while saving at least 50k update steps across the majority of metrics. 
This efficiency is particularly evident for the CLAP Score and KL metrics, which achieve parity up to 190k training steps earlier.

\subsection{Ablation}

Here we provide ablation studies, in which we evaluate the two individual semantic and acoustic features as well as the proposed unified feature for understanding and reconstruction tasks.
For understanding tasks, we directly evaluate each individual feature $z_{\text{sem}}, z_{\text{ac}}, z$.
For reconstruction, we utilize the decoder $G$ to synthesize the waveform directly from the respective features.
The results in \Cref{tab:understanding_ablation} indicate that the semantic features provides the primary signal for most classification tasks, while the acoustic features alone yields poor performance. 
However, the unified feature significantly can enhance performance in paralinguistic (CREMA-D) and music tasks (FMA, NSynth), demonstrating that injected acoustics can complement semantic features for audio classification.

\begin{table}[htb]
    \centering
    \caption{Audio understanding performance on the X-ARES benchmark for the individual acoustic, semantic and proposed unified features. Higher is better and best is in \textbf{bold}.}
    \label{tab:understanding_ablation}
    \begin{tabular}{l|rccc}
\toprule
Domain & Task & Semantic & Acoustic & {Unified} \\
\midrule
\multirow{11}{*}{Speech} & CREMA-D & 80.10  & {44.44} & \textbf{80.55} \\
 & FSC & \textbf{98.26} & 2.71 & 83.39    \\
 & LibCnt & \textbf{80.79} &  52.34 & {79.98}  \\
 & LS100h & \textbf{80.32} & 0.00 & 68.70 \\
 & LSMF & \textbf{98.47} & 88.14 & 98.06\\
 & RAV & \textbf{87.15} &  33.54 & 83.06 \\
 & SPV1 & \textbf{97.61} & 19.62 & 97.35  \\
 & VocI & \textbf{30.32} & 2.86 & 23.92 \\
 & VocS & \textbf{93.15} & 47.73 & \textbf{93.15} \\
 & VoxC & \textbf{55.51} & 19.62 & 51.16 \\
 & VoxL33 & \textbf{75.40} & 34.34 & 74.06  \\
\midrule
\multirow{7}{*}{Sound} & ASV2015 & \textbf{98.37} & 89.73 & 96.45 \\
 & Clo & \textbf{6.12} & 0.37 & 5.63 \\
 & DES & 52.02 & 21.49 & \textbf{55.40} \\
 & ESC & \textbf{96.95} & 34.05 & 96.40 \\
 & FSD50 & \textbf{61.46} & 8.93 & 59.98 \\
 & F18K & \textbf{89.19} & 27.06 & 86.81 \\
 & Urb8 & \textbf{86.23} & 50.02 & 83.80 \\
\midrule
\multirow{4}{*}{Music} & FMA & 65.14 & 41.71 & \textbf{66.29} \\
 & GTZAN & \textbf{91.19} & 58.85 & 89.99 \\
 & MAESTRO & 38.65 & \textbf{58.84} & {57.65} \\
 & NSynth & 77.59 & 43.16 & \textbf{77.91} \\
\bottomrule
    \end{tabular}
\end{table}
We further ablate the importance of the injected acoustic features in DashengTokenizer, where we evaluate the reconstruction performance of each feature individually.
The results in \Cref{tab:ablation_reconstruction} unsurprisingly show that the semantic features alone cannot be used for effective reconstruction.
However, our proposed unified feature still achieves similar performance to the acoustic feature, making it suitable for both understanding and generation tasks.

Overall, these results demonstrate that our unified approach successfully bridges the gap between maintaining high-fidelity acoustic information without compromising on general-purpose audio understanding.
\begin{table}[htb]
\centering
\caption{Reconstruction performance for semantic, acoustic and the proposed unified features. Best are in \textbf{bold}.}
\label{tab:ablation_reconstruction}
\begin{tabular}{lcccccc}
\toprule
\multirow{2}{*}{Feature} & \multicolumn{2}{c}{Audioset} & \multicolumn{2}{c}{MUSDB18} & {SEED-ZH} & SEED-EN  \\
\cmidrule(lr){2-3} \cmidrule(lr){4-5} \cmidrule(lr){6-7}
& Mel-16k $\downarrow$ & STFT-16k  $\downarrow$ & Mel-16k $\downarrow$ & STFT-16k  $\downarrow$ & \multicolumn{2}{c}{PESQ $\uparrow$} \\
\midrule
Acoustic            & \textbf{0.268} & \textbf{1.497} & \textbf{0.267} & \textbf{1.410} & \textbf{4.178} & \textbf{4.131} \\
Semantic            & 3.310 & 4.978  & 3.290  & 5.540 & 1.040 & 1.040 \\
{Unified}            & {0.320} & {1.585} & {0.293} & {1.458} & 4.163  & 4.125  \\
\bottomrule
\end{tabular}
\end{table}

\section{Conclusion}

This paper proposed DashengTokenizer, a unified audio embedding that can be used for generation and understanding tasks.
Our proposed approach trains a simple linear projection to map acoustic details into high-level semantic features.
The resulting approach results in a competitive performance in terms of reconstruction quality, while also outperforming audio encoders and codecs on a plethora of understanding tasks.
Further, DashengTokenizer is then applied for SE tasks, where it is shown that this tokenizer achieves superior performance compared to baselines in SE.
Lastly, we show that using DashengTokenizer, training efficiency for TTA and TTM tasks largely outperform competitive VAE baselines.

\clearpage
\bibliographystyle{plain}
\bibliography{papers} 


\appendix

\newpage

\appendix

\section{Ablation experiments for semantic weight}
\label{sec:ablation_weight}

We evaluate the impact of the semantic loss weight ($\lambda_{\text{sem}}$) on the model's performance. 
To investigate the trade-off between semantic representation and acoustic fidelity, we perform a parameter sweep by varying the semantic weight $\lambda_{\text{sem}}$ from 10 to 60 while keeping the mel-reconstruction weight $\lambda_{\text{mel}}$ fixed at 45. 
We further include two boundary cases: $\lambda_{\text{sem}}=0$ (purely acoustic features) and $\lambda_{\text{mel}}=0$ (purely semantic features) to establish performance baselines/toplines for each objective.
We assess the model across two dimensions: first, audio reconstruction performance using Mel-16k distance, and second, the quality of the learned representations for downstream classification tasks across diverse audio domains.
For all the following experiments we use a 86M parameter encoder, which is pretrained on the same training dataset as MiDashengLM-7B. 
Training for this ablation is done for 200k iterations solely on the LibriSpeech dataset.

\begin{table}[ht]
\centering
\begin{tabular}{rr|ccc}
\toprule
$\lambda_{\text{sem}}$ & $\lambda_{\text{mel}}$ & LibriTTS-clean & Audioset & MusicCaps \\
\cmidrule(lr){3-5}
& &  \multicolumn{3}{c}{Mel-16k $\downarrow$}\\
\midrule
0 & 45 &  0.33 & 0.37 & 0.41  \\
\midrule
10 & 45 & 0.35 & 0.44 & 0.52 \\
30 & 45 & 0.35 & 0.47 & 0.56 \\

\rowcolor{xiaomimedgray!10} 45 & 45 & 0.35 & 0.49 & 0.59 \\
60 & 45 & 0.48 & 0.61 & 0.68 \\
\midrule
45 & 0 &  1.47 & 4.15 & 4.96  \\
\bottomrule
\end{tabular}
\caption{Reconstruction Performance metrics for different $\lambda_{\text{sem}},\lambda_{\text{mel}}$. Configuration used in this paper is highlighted.}
\label{tab:audio_reconstruct_ablate}
\end{table}

\begin{table}[ht]
\centering
\begin{tabular}{rr|ccc}
\toprule
$\lambda_{\text{sem}}$ & $\lambda_{\text{mel}}$ & LibriTTS-clean & Audioset & MusicCaps \\
\midrule
0 & 45 & \textcolor{xiaomired}{-6\%} & \textcolor{xiaomired}{-24\%} & \textcolor{xiaomired}{-31\%} \\
\midrule
10 & 45 & 0\% & \textcolor{xiaomired}{-10\%} & \textcolor{xiaomired}{-12\%} \\
30 & 45 & 0\% & \textcolor{xiaomired}{-4\%} & \textcolor{xiaomired}{-5\%} \\
\rowcolor{xiaomimedgray!10} 45 & 45 & 0.35 & 0.49 & 0.59 \\
60 & 45 & \textcolor{xiaomiblue}{+37\%} & \textcolor{xiaomiblue}{+24\%} & \textcolor{xiaomiblue}{+15\%} \\
\midrule
45 & 0 & \textcolor{xiaomiblue}{+320\%} & \textcolor{xiaomiblue}{+747\%} & \textcolor{xiaomiblue}{+741\%} \\
\bottomrule
\end{tabular}
\caption{Relative reconstruction performance change ($\Delta$ vs. $\lambda_{\text{sem}} = 45, \lambda_{\text{mel} = 45}$). Configuration used in this paper is highlighted.}
\label{tab:audio_reconstruct_delta}
\end{table}

The results in \Cref{tab:audio_reconstruct_ablate} indicate a clear trade-off between semantic constraint ($\lambda_{\text{sem}}$) and acoustic fidelity. 
As the weight of the semantic loss increases from 0 to 60, we observe a consistent rise in the Mel-16k reconstruction error across all three datasets. 
Most notably, when only reconstructing from semantic features and setting $\lambda_{\text{mel}}=0$, we can observe a significant increase in Mel-16k Distance on the Audioset and MusicCaps datasets.

Interestingly, these results highlight a significant divide in how acoustic and semantic features generalize across domains. 
When relying strictly on acoustic features ($\lambda_{\text{sem}} = 0, \lambda_{\text{mel}} = 45$), the model maintains strong performance even on out-of-domain data like MusicCaps, implying that low-level acoustic representations are inherently more universal.
However, when using purely semantic features for reconstruction ($\lambda_{\text{sem}} = 45, \lambda_{\text{mel}} = 0$) the performance degradation on non-speech datasets is more than double that of the clean speech in LibriTTS.
This shows that semantic features are inherently task specific and do not generalize across domains.

As shown in \Cref{tab:audio_reconstruct_delta}, lower values of $\lambda_{\text{sem}}$ facilitate better reconstruction, providing up to a 31\% reduction in Mel-distance compared to the $\lambda_{\text{sem}} = 45$ baseline. 
In contrast, increasing $\lambda_{\text{sem}}$ to 60 or removing the Mel-reconstruction loss entirely ($\lambda_{\text{mel}} = 0$) leads to substantial performance drops, with the latter causing error increases of over 700\% in non-speech domains.

\begin{table}[ht]
\centering
\resizebox{\textwidth}{!}{%
\begin{tabular}{lcccccccccccc}
\toprule
$\lambda_{\text{sem}}$ & $\lambda_{\text{mel}}$ & ESC & FSD50 & GTZAN & LS100h & NSynth & SPV1 & Urb8 & VocS & VoxC & VoxL33 \\
\midrule
0 & 45 & 35.20 & 8.20 & 54.16 & 0.00 & 38.67 & 24.77 & 51.97 & 50.29 & 21.62 & 10.07 \\
\midrule
10 & 45 & 85.25 & 53.11 & 81.78 & 78.43 & 76.56 & 97.23 & 79.11 & 92.99 & 30.94 & 90.30 \\
30 & 45 &  90.25& 55.26&	85.89&	80.23 &	76.90 &	97.29 &	81.18 &	93.13 &	33.67 &	90.55 \\
45 & 45 & \textbf{91.85} & 55.94 & \textbf{86.89} & 81.89 & 77.05 & 97.29 & 81.83 & \textbf{93.15} & \textbf{37.20} & 91.11 \\
60 & 45 & 91.60 & \textbf{56.09} & 86.88 & 85.39 & \textbf{77.17} & \textbf{97.40} & 
\textbf{82.12} & 92.79 & 35.40 & \textbf{91.30} \\
\midrule
45 & 0 & 95.35&	60.43	& 89.09	& 83.67	&77.86	&97.57	&85.49&	93.71	&32.61&	91.86 \\ 
\bottomrule
\end{tabular}%
}
\caption{Downstream understanding performance for different $\lambda_{\text{sem}},\lambda_{\text{mel}}$.}
\label{tab:audio_understanding_ablate}
\end{table}

\begin{table}[htbp]
\centering
\caption{Relative performance change in percent for downstream tasks (Ablation $\Delta$ vs. $\lambda_{\text{sem}} = 45,\lambda_{\text{mel}}=45$).}
\label{tab:audio_benchmarks_delta}
\resizebox{\textwidth}{!}{%
\begin{tabular}{lcccccccccccc}
\toprule
$\lambda_{\text{sem}}$ & $\lambda_{\text{mel}}$ & ESC & FSD50 & GTZAN & LS100h & NSynth & SPV1 & Urb8 & VocS & VoxC & VoxL33 \\
\midrule
0 & 45 & \textcolor{xiaomiblue}{-62\%} & \textcolor{xiaomiblue}{-85\%} & \textcolor{xiaomiblue}
{-38\%} & \textcolor{xiaomiblue}{-100\%} & \textcolor{xiaomiblue}{-50\%} & \textcolor{xiaomiblue}{-75\%} & \textcolor{xiaomiblue}{-36\%} & \textcolor{xiaomiblue}{-46\%} & \textcolor{xiaomiblue}{-42\%} & \textcolor{xiaomiblue}{-89\%} \\
\midrule
10 & 45 & \textcolor{xiaomiblue}{-7\%} & \textcolor{xiaomiblue}{-3\%} & \textcolor{xiaomiblue}{-5\%} & \textcolor{xiaomiblue}{-3\%} & \textcolor{xiaomiblue}{-0.5\%} & \textcolor{xiaomiblue}{-0.1\%} & \textcolor{xiaomiblue}{-3\%} & \textcolor{xiaomiblue}{-0.2\%} & \textcolor{xiaomiblue}{-6\%} & \textcolor{xiaomiblue}{-0.8\%} \\
30 & 45 & \textcolor{xiaomiblue}{-2\%} & \textcolor{xiaomiblue}{-1\%} & \textcolor{xiaomiblue}{-1\%} & \textcolor{xiaomiblue}{-2\%} & \textcolor{xiaomiblue}{-0.2\%} & 0.0\% & \textcolor{xiaomiblue}{-0.8\%} & 0.0\% & \textcolor{xiaomiblue}{-9\%} & \textcolor{xiaomiblue}{-0.6\%} \\ 
45 & 45 & 0.0\% & 0.0\% & 0.0\% & 0.0\% & 0.0\% & 0.0\% & 0.0\% & 0.0\% & 0.0\% & 0.0\% \\
60 & 45 & \textcolor{xiaomiblue}{-0.3\%} & \textcolor{xiaomired}{+0.2\%} & 0.0\% & \textcolor{xiaomired}{+4\%} & \textcolor{xiaomired}{+0.1\%} & \textcolor{xiaomired}{+0.1\%} & \textcolor{xiaomired}{+0.3\%} & \textcolor{xiaomiblue}{-0.4\%} & \textcolor{xiaomiblue}{-2\%} & \textcolor{xiaomired}{+0.2\%} \\
\midrule
45 & 0 & \textcolor{xiaomired}{+4\%} & \textcolor{xiaomired}{+8\%} & \textcolor{xiaomired}{+3\%} & \textcolor{xiaomired}{+2\%} & \textcolor{xiaomired}{+1\%} & \textcolor{xiaomired}{+0.3\%} & \textcolor{xiaomired}{+4\%} & \textcolor{xiaomired}{+0.6\%} & \textcolor{xiaomiblue}{-12\%} & \textcolor{xiaomired}{+0.8\%} \\
\bottomrule
\end{tabular}
}
\end{table}

Results for audio understanding can be seen in \ref{tab:audio_understanding_ablate}, where we see that performance scales sharply as $\lambda_{\text{sem}}$ increases from 0 to 45. 
However, further increasing the weight to $\lambda_{\text{sem}} = 60$ yields marginal gains (e.g., $+4\%$ on LS100h, $+0.2\%$ on FSD50).
These slight gains for semantic understanding come at the price of significant, up to 37\%, performance drops in terms of reconstruction performance (see \Cref{tab:audio_benchmarks_delta}). 

Consequently, we identify $\lambda_{\text{sem}} = 45$ as the optimal balance between semantic  and reconstruction performance.

\section*{Acknowledgement}

This work makes use of the Million Song Dataset, the MTG-Jamendo~\cite{bogdanov2019mtg} Dataset, AudioCaps~\cite{kim2019audiocaps}, Libri-light~\cite{kahn2020libri}, AudioSet~\cite{gemmeke2017audioset}, ACAVCaps~\cite{niu2025mecat,niu2026acavcapsenablinglargescaletraining} and AIShell-1/2/3~\cite{bu2017aishell,shi2020aishell3,du2018aishell2} datasets. 
The authors confirm that the use of the MTG-Jamendo, AudioCaps, and AudioSet datasets is strictly limited to academic research purposes and does not involve any commercial activities. 
Furthermore, in accordance with the license terms of Libri-light, we confirm its application is restricted to model evaluation. All datasets are used in compliance with their respective licensing agreements and original citations.

\end{document}